\documentclass[prb,twocolumn,groupedaddress,amsfont,showpacs]{revtex4}
\usepackage{amssymb}

\usepackage{graphicx}

\newcommand{\rads}[0]{rad s$^{-1}$}

\begin{document}

\title{Using acoustic waves to induce high-frequency current oscillations in superlattices}

\author{M.T. Greenaway$^1$,  A.G. Balanov$^{1,2}$, D. Fowler$^1$, A.J. Kent$^1$, and T.M. Fromhold$^1$} 
\affiliation{$^1$School of Physics and Astronomy, University of Nottingham, Nottingham NG7 2RD, United Kingdom \\
$^2$Department of Physics, Loughborough University, Leicestershire, LE11 3TU, United Kingdom}
\date{\today}

\begin{abstract}
We show that GHz acoustic waves in semiconductor superlattices can induce THz electron dynamics that depend critically on the wave amplitude. Below a threshold amplitude, the acoustic wave drags electrons through the superlattice with a peak drift velocity overshooting that produced by a static electric field. In this regime, single electrons perform drifting orbits with THz frequency components. When the wave amplitude exceeds the critical threshold, an abrupt onset of Bloch-like oscillations causes negative differential velocity. The acoustic wave also affects the collective behavior of the electrons by causing the formation of localised electron accumulation and depletion regions, which propagate through the superlattice, thereby producing self-sustained current oscillations even for very small wave amplitudes. We show that the underlying single-electron dynamics, in particular the transition between the acoustic wave dragging and Bloch oscillation regimes, strongly influence the spatial distribution of the electrons and the form of the current oscillations. In particular, the amplitude of the current oscillations depends non-monotonically on the strength of the acoustic wave, reflecting the variation of the single-electron drift velocity.
\end{abstract}



\maketitle

\section{Introduction}
Electrons in semiconductor superlattices (SLs) exhibit a wide range of nonlinear effects that are of fundamental scientific interest and useful for applications in ultrafast electronics \cite{esaki-tsu,dohler,IGN76,mendez,sibille,Holthaus,sakaki,KEA95,leb,canali,ZHA96,ALE96,schom,WAC02,AMA02a,Pat02,FRO04,SOSKIN,ALE07,NU2299,Renk,Fowler,Balanov}. Many of these effects originate from the SL minibands, which enable electrons to perform THz frequency Bloch oscillations when a sufficiently high static electric field is applied along the SL axis \cite{esaki-tsu,dohler,sibille,WAC02}. Bloch oscillations cause the electron drift velocity to decrease with increasing electric field, which can trigger charge-domain oscillations accompanied by the emission of electromagnetic radiation \cite{ZHA96,schom}. 

The frequency response of SL oscillators and detectors is limited by scattering processes including electron-phonon interactions \cite{sakaki,schom,WAC02}. Surprisingly, though, phonons can serve as a powerful tool for \emph{enhancing} the electronic and optical properties of solid state devices \cite{nature,ahn}. For example, in `SASER' SLs \cite{kent_saser}, analogous to the laser, the amplification of coherent sound waves now opens the way to acoustic control of miniband electron transport. In related work, we recently demonstrated that a train of acoustic strain pulses can induce current in a SL by dragging miniband electrons through the device \cite{FOW2008}. 

In this paper, we show that a continuous GHz acoustic wave can create complex THz electron dynamics in SLs, thus producing high-frequency current oscillations even when no static electric field is applied. In a single-electron picture, there are two distinct dynamical regimes, depending on whether the energy amplitude, $U$, of the acoustic wave is greater, or less, than a critical value, $U_c$, which depends on the SL parameters. For $U<U_c$, the acoustic wave drags electrons through the SL, producing, in the presence of electron scattering, a drift velocity, $v_d$, far higher than the speed of the wave itself. In this regime, the electrons perform periodic orbits in the rest frame of the acoustic wave. The orbital frequencies are more than an order of magnitude higher than the sound wave frequency. When $U>U_c$, the acoustic wave no longer drags electrons through the SL. Instead, there is an abrupt onset of Bloch-oscillation bursts, 
which makes $v_d$ decrease extremely rapidly with increasing $U$. 

Our results demonstrate that miniband electrons driven by a GHz acoustic wave can attain a higher maximal $v_d$, and have a larger negative differential velocity (NDV), than when they are accelerated by a \emph{static} electric field. Since high-frequency SL oscillators require high values of both parameters \cite{schom,WAC02}, acoustic wave driving could strongly enhance the performance of such devices. To investigate how the sudden onset of single-particle Bloch oscillations affects the collective dynamics of the electrons, we solve the Poisson and (drift-diffusion) current-continuity equations self-consistently throughout the SL. We use our results to determine the current through the SL as a function of $U$ and time. Remarkably, the current exhibits self-sustained high-frequency oscillations for all $U$. For $U < U_c$, these oscillations originate from space-charge density waves, created and dragged through the SL by the acoustic wave. In this regime, as $U$ increases the space-charge modulation strengthens, and so the amplitude of the current oscillations also increases. But when $U$ increases above $U_c$, the onset of Bloch oscillations localises the electrons, thereby weakening the current oscillations. This electron localisation also leads to the formation of propagating charge domains, which coexist with the charge density waves. 

The complex single-particle and collective electron dynamics induced by an acoustic wave demonstrate the potential of SLs for interfacing high-frequency electronics with the emerging field of phononics \cite{CHA06,WANG07}, which is attracting considerable interest in applied physics and engineering. In particular, our results indicate that SLs can be used to both detect and up-convert an applied acoustic signal. Generic features of the energy band transport processes created by a propagating wave potential also suggest ways to control transport through other spatially-periodic systems, including cold atoms in optical lattices \cite{renzoni,kos}.   

The structure of the paper is as follows. In Section \ref{sec:Sec1}, we introduce the semiclassical equations of motion for a miniband electron in the presence of an acoustic wave and show that the electron dynamics and drift velocity depend critically on the wave amplitude. In Section \ref{sec:Sec2}, we consider the collective electron dynamics and show that the acoustic wave triggers linear charge density waves, nonlinear charge domains (when $U>U_c$), and self-sustained current oscillations even for very small wave amplitudes. Finally, in Section \ref{sec:Sec3}, we summarise our results and draw conclusions.

\begin{figure}
  \centering
\includegraphics*[width=1.\linewidth]{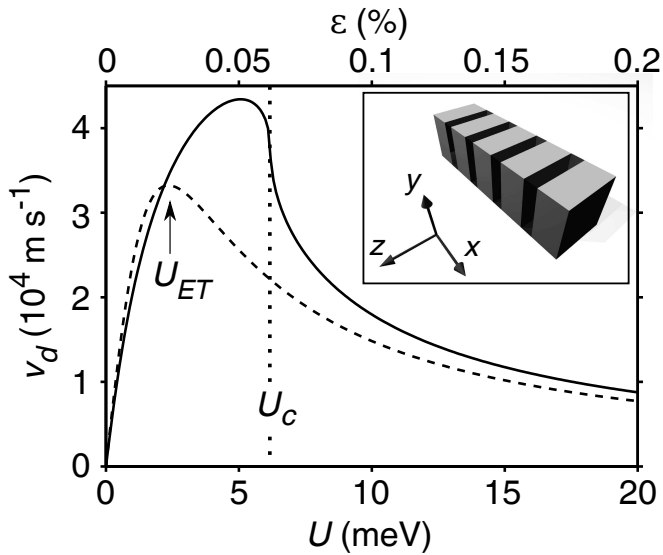}
  \caption{Solid [dashed] curve: $v_d$ versus $U$ (lower scale) or $\varepsilon$ (upper scale) calculated for a miniband electron driven by an acoustic wave only [or accelerated by a \emph{constant} electric field, $k_SU/e$, only]. Dotted line [arrow] marks $U= U_c  [U_{ET}]$. Inset: schematic diagram of the SL layers and co-ordinate axes.
\label{fig:vdenergy}}
\end{figure}

\section{Model of single electron dynamics}
\label{sec:Sec1} 

We consider a longitudinal acoustic wave, which propagates along the SL $x$ - axis (Fig. 1 inset), creating a position and time ($t$) dependent potential energy field, $V(x,t)= -U \sin \left( k_S x - \omega_S t \right)$, for each miniband electron \cite{foot2,kent_metalfilm}. 
The wave amplitude, $U=\varepsilon D$, depends on the maximum strain, $\varepsilon < 0.5\%$, that the acoustic wave creates and on the deformation potential, $D$ \cite{phononspec}. We consider acoustic waves whose wavenumber, $k_S$, lies within the inner half of the minizone, so that there is linear frequency dispersion $\omega_S=v_S k_S$, where $v_S$ is the speed of sound. Since the sound wave exerts force along $x$ only, the electron dynamics can be described by a one-dimensional model for motion in the lowest miniband. Within the tight-binding approximation, the kinetic energy versus crystal momentum dispersion relation for this miniband is $E(p_x) =\Delta[1-\cos (p_x d / \hbar)]/2$, where $\Delta$ is the miniband width, and $d$ is the SL period \cite{WAC02}. We take $\Delta = 7$ meV, $d = 12.5$ nm, $D = 10$ eV, and $v_S = 5000$ m s$^{-1}$, corresponding to a GaAs/(AlGa)As SL used in recent experiments \cite{NU2299,phononspec}, but obtain similar results for a wide range of SL parameters. The wave is sufficiently weak and spatially slowly varying to preserve the miniband \cite{WAC02}, thus ensuring the validity of a semiclassical model \cite{footac}.

The semiclassical equations of electron motion are 
\begin{eqnarray}
v_x=\frac{dx}{dt}=\frac{\partial H}{\partial p_x}=\frac{\Delta d}{2\hbar} \sin \left(\frac{p_x d}{\hbar} \right),  \label{eq:simx} \\
\frac{dp_x}{dt}=-\frac{\partial H}{\partial x}=k_S U\cos(k_S (x+x_0) - \omega_S t), \label{eq:pxdotdef}
\end{eqnarray}
where the Hamiltonian $H(x,p_x)=E(p_x)+V(x,t)$. We solve Eqs. (1,2) numerically, taking $v_x=0$, and $p_x$=0 when $t=0$, to determine the electron trajectories in the absence of scattering. 


\subsection{Electron dynamics for initial position $x_0=0$}

In order to understand the general dynamics of a single electron, first we consider the simplest situation by setting $x_0=x(t=0)=0$.  

We use the Esaki-Tsu model \cite{esaki-tsu,footnoteC} to find the electron drift velocity 

\begin{equation}
v_d=\langle v_x (t)\exp(-t/\tau)\rangle/\tau,
\label{eq:driftvelocity}
\end{equation}
where $\langle .\rangle$ denotes integration over $t>0$, taking, from experiment \cite{NU2299}, an electron scattering time $\tau=250$ fs, which includes both elastic (interface roughness) and inelastic (phonon) scattering.

The solid curve in Fig. \ref{fig:vdenergy} shows $v_d$ calculated as a function of $U$ (lower scale), or, equivalently, $\varepsilon$ (upper scale) for an acoustic wave with $\omega_S = 4\times10^{11}$ rad s$^{-1}$ and wavelength $\lambda_S = 2 \pi/k_S \approx 6d$ less than the length of most SLs \cite{footnoteF,footnoteD,FRO04,SOSKIN}. For comparison, the dashed curve shows the usual Esaki-Tsu drift velocity \cite{esaki-tsu} calculated for an electron accelerated by a \emph{static} electric field of magnitude $k_SU/e$, where $e$ is the magnitude of the electronic charge. As discussed extensively in the literature \cite{esaki-tsu,dohler,WAC02}, the Esaki-Tsu $v_d(U)$ curve is linear for small $U$, attains a maximum when $U=U_{ET}=\hbar/\tau\approx2.4$ meV, and thereafter decreases with increasing $U$ as more electrons complete Bloch oscillations before scattering.
Figure \ref{fig:vdenergy} reveals that for both low and high $U$, the $v_d(U)$ curves produced by the static force and acoustic wave converge. But for intermediate $U$ there are major differences in the two curves. In particular, the acoustic wave generates a larger peak $v_d$ value and a far steeper (factor $\approx13$) NDV region.
\begin{figure}
  \centering
\includegraphics*[width=1.\linewidth]{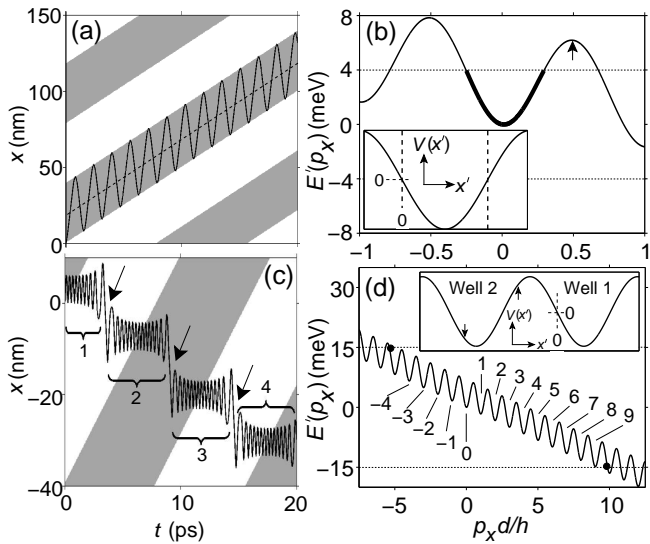}
  \caption{(a) Solid curve: electron trajectory, $x(t)$, calculated for $U=$ 4 meV. Within the white [gray] regions, $V(x,t)$ is $>0$ [$<0$]. Dashed line has gradient $v_S$. (b) $E'(p_x)$: dotted lines mark $\pm U$ when $U=4$ meV. At arrowed peak, $E'(p_x)=U_c$. Inset: $V(x')$, where dotted lines mark turning points of orbit in (a). (c) As (a) except $U=$ 15 meV. Bloch oscillation bursts, within numbered brackets, are separated by sudden jumps (arrowed). (d) As (b), except $U=15$ meV. Left- [right-] hand filled circles mark where $E'(p_x)$ = $U$ [-$U$]. Numbers label different minizones. Inset: adjacent wells (1 and 2) in $V(x')$, with arrows discussed in text.
\label{fig:trajs}}
\end{figure}

To explain these differences, we  consider the electron dynamics in the absence of scattering. Figure \ref{fig:trajs}(a) shows the $x(t)$ trajectory obtained numerically from Eqs. (1,2), taking $U=4$ meV, below the peak in the $v_d(U)$ curve (solid curve in Fig. \ref{fig:vdenergy}) generated by the acoustic wave. The trajectory consists of regular, almost sinusoidal, oscillations superimposed on a linear background of gradient $v_S$ [dashed line in Fig. \ref{fig:trajs}(a)], suggesting that the acoustic wave drags the electron through the SL \cite{FRO93,FRO93b}. We confirm this picture by considering electron motion in the \emph{rest frame} of the acoustic wave, in which the electron's position, $x'(t)=x(t)-v_St$, determines the \emph{static} potential energy $V(x')= -U \sin (k_S x')$. In this frame, the Hamiltonian is $H'(x',p_x)=E'(p_x)+V(x')$, where $E'(p_x)=E(p_x)-v_Sp_x$, and the equations of motion are
\begin{eqnarray}
v'_x=v_x-v_S=\frac{dx'}{dt}=\frac{\partial H'}{\partial p_x}=\frac{\Delta d}{2\hbar} \sin \left(\frac{p_x d}{\hbar} \right)-v_S \label{eq:simx2}, \\
\frac{dp_x}{dt}=-\frac{\partial H'}{\partial x'}=k_SU\cos(k_Sx'). \label{eq:pxdotdef2}
\end{eqnarray}

Since $H'$ is not an explicit function of $t$, it is a constant of the motion but does \emph{not} equal the total energy, $H$. For the initial conditions considered here, $H'=0$, meaning that $E'(p_x)=-V(x')$ can only take values between $\pm U$, marked by the horizontal dotted lines in Fig. 2(b) for the trajectory in Fig. 2(a). The lines reveal that the electron can only access the almost parabolic region of the $E'(p_x)$ curve [thick in Fig. 2(b)] around $p_x=0$. Since for the given parameters the minimum value of $E'(p_x)$ that the electron can attain is close to zero, its maximum potential energy is also close to zero. The electron is therefore confined within a single potential well in the acoustic wave and oscillates back and forth across this well between turning points at $x'=0$ and $\lambda_S/2$ [vertical dashed lines in Fig. 2(b) inset]. Since the electron remains within the almost parabolic region of $E'(p_x)$, where its effective mass is constant, $x'(t)$ is an almost harmonic function of $t$. Therefore we can approximate $x(t)$ as 

\begin{equation}
x(t) \approx v_St + \lambda _S[1-\cos\left(\omega_Rt\right)]/4,
\label{eq:xt}
\end{equation}

\noindent
where $\omega_R$ is the frequency for motion to and fro across the potential well. This 
approximation accurately describes electron trajectories for small $U$, for example that shown in Fig. 2(a). The electron is trapped in the well, where $V(x,t)\lesssim 0$ [gray bands in Fig. 2(a)]. But as the well moves, it drags the electron through the SL with a mean speed equal to $v_S$ \emph{in the absence of scattering}. 

Increasing $U$ above 4 meV initially has no qualitative effect on the electron orbits. They continue to be dragged through the SL and are of the form $x(t)=v_St+f(t)$, where the periodic function, $f(t)$, becomes less harmonic as $U$ [upper dotted line in Fig. 2(b)] increases, thus making the electron access nonparabolic regions of $E'(p_x)$.

When $U$ reaches a critical value, $U_c$, equal to the local maximum of $E'(p_x)$ marked by the arrow in Fig. 2(b), the electron trajectories can reach the edge of the first minizone, and therefore change abruptly from closed to open orbits that can traverse several minizones. The local maximum of $E'(p_x)$ occurs when $dE(p_x)/dp_x=v_S$, i.e. when $\sin (p_x d/ \hbar)= 2\hbar v_S/ \Delta d$ ($\approx 0.08$ for the given parameters). Using small-angle approximations, it follows that the local maximum occurs when $p_x  \approx ( \hbar \pi/d)-(2 \hbar^{2} v_S/ \Delta \ d^{2})=p_x^m$. Therefore, from  $E'(p_x^m)=U_c$, we estimate  

\begin{equation}
U_c \approx \Delta -v_S \hbar \pi/d.
\label{eq:uc}
\end{equation}

Figure \ref{fig:trajs}(c) shows $x(t)$ calculated for $U=15$ meV $>U_c$ = 6.2 meV. The bursts of high-frequency fluctuations in $x(t)$ (within brackets) are Bloch oscillations driven by the acoustic wave. The jumps in $x(t)$ (arrowed) occur at the centers of the white and gray stripes in Fig. 2(c), when $V$ is extremal  and, consequently, the acoustic force is zero, and therefore unable to induce Bloch oscillations.

To explain fully the form of the trajectory in Fig. 2(c), we consider the electron motion in the rest frame of the acoustic wave. Initially, the electron is at $x'=0$ where the high gradient of $V(x')$ [Fig. \ref{fig:trajs}(d) inset] causes $p_x$ rapidly to increase to the edge of the first minizone [labeled 0 in Fig. 2(d)], thus reversing $v_x$ and $v'_x$. After crossing the minizone boundary, the electron continues to experience a large positive force, which increases $p_x$ through minizones 1-9 in Fig. 2(d), thus generating the Bloch oscillations within Bracket 1 in Fig. 2(c). As $p_x$ increases, the average value of $E'(p_x)$ decreases [Fig. 2(d)] and $V(x')$ increases (to keep $H'=0$) as the electron moves up the left-hand side of Well 1 in Fig. \ref{fig:trajs}(d) inset. As the electron climbs the well wall, $\vert dV(x')/dx' \vert$ decreases, thus reducing the frequency of the Bloch oscillations and increasing their amplitude \cite{esaki-tsu,dohler,WAC02}, as shown by the $x(t)$ curve within Bracket 1 in Fig. 2(c). 

When the electron reaches the top of Well 1, so that $V(x')=U$, $E'(p_x)$ attains its lowest possible value of $-U$ [lower dotted curve in Fig. 2(d)] and so $p_x$ can no longer increase. Instead, since the acoustic force is instantaneously zero, $p_x$ is temporarily pinned at the intersection [right-hand filled circle in Fig. 2(d)] between $E'(p_x)$ and the lower dotted line. The large negative velocity at this intersection, $dE'/dp_x\approx -5.6\times10^{4}$ m s$^{-1}$, makes the electron jump backwards along the section of the $x(t)$ curve marked by the left-hand arrow in Fig. 2(c). This jump transfers the electron to the position marked by the right-hand arrow in Well 2 [Fig. \ref{fig:trajs}(d) inset]. At this position, the acoustic wave exerts a large negative force on the electron, which causes $p_x$ to decrease, so inducing another burst of Bloch oscillations [within Bracket 2 in Fig. 2(c)], until $E'(p_x)$ reaches its maximum value [upper dotted line in Fig. 2(d)] and $V(x')$ attains its minimum value of $-U$ in Well 2. Then, the electron again jumps backwards, along the $x(t)$ trajectory marked by the central arrow in Fig. 2(c), with velocity $\approx -6.8\times10^{4}$ m s$^{-1}$, approximately equal to $dE'/dp_x$ at the intersection [left-hand filled circle in Fig. 2(d)] between $E'(p_x)$ and the upper dotted line. This jump transfers the electron to the position marked by the left-hand arrow in Well 2 [Fig. \ref{fig:trajs}(d) inset], where a large positive force causes $p_x$ rapidly to increase, triggering the Bloch oscillation burst within Bracket 3 in Fig. 2(c). Thereafter, the cycle repeats, with the electron jumping backwards after each Bloch oscillation burst. 

The number of Bloch oscillations within each burst equals the number of distinct minizones, $N\approx 2Ud/h v_S$, that the electron traverses. When $U=$ 15 meV, $N=14$, corresponding to crossing the minizones labeled -4 to 9 in Fig. 2(d). The abrupt onset of the acoustically-driven Bloch oscillations contrasts with the gradual switch on produced by increasing a \emph{static} force \cite{esaki-tsu,dohler,WAC02}. 
\begin{figure}
  \centering
\includegraphics*[width=1.\linewidth]{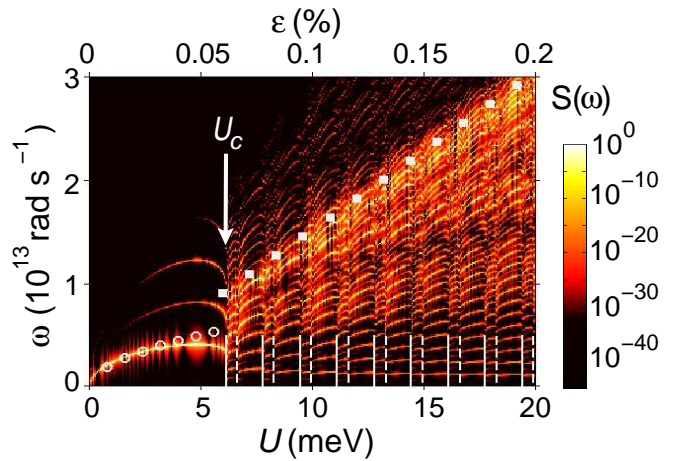}
  \caption{(Color online) Fourier power spectrum, $S(\omega)$, (scale right) of $v_x(t)$ trajectories calculated for a range of $U$ (lower scale) or $\varepsilon$ (upper scale) at fixed $\omega_S=4\times10^{11}$ rad s$^{-1}$. Arrow is at $U=U_c$. Open circles [filled squares]: analytical estimates of $\omega_R$ [$\omega_{B}^{max}$] in the wave-dragging [Bloch oscillation] regimes. Solid and dashed white lines are discussed in text.
\label{fig:spec}}
\end{figure}

Figure \ref{fig:spec} shows a color map of the Fourier power, $S(\omega)$, of $v_x(t)$ trajectories calculated for a range of $U$ at fixed $\omega_S=4\times10^{11}$ rad s$^{-1}$. The spectrum changes abruptly at $U=U_c$ (arrowed), due to the transition from the wave-dragging to Bloch oscillation regimes. 

For $U<U_c$, $S(\omega)$ has a sharp peak (lower left light curve in Fig. \ref{fig:spec}) at the frequency, $\omega_R$, for motion across the potential well [Fig. 2(b) inset] that traps the electron and drags it through the SL. Three higher harmonics are also visible in the color map, but their power is orders of magnitude lower than the fundamental. 
When $U\approx4$ meV, $\omega_R\approx17 \omega_S$, indicating that the dragged electron paths cause significant frequency up-conversion of the acoustic wave. 
In the regime $U\lesssim U_c$, corresponding to periodic $x'(t)$ trajectories, the equations of motion yield a simple equation for $\omega_R \approx \alpha \omega_S$, where the factor $\alpha = (U\Delta/\pi)^{1/2}(d/\hbar v_S)$ is estimated by substitution of the approximation (\ref{eq:xt}) into the set of equations (\ref{eq:simx2}) and (\ref{eq:pxdotdef2}). This factor can be used to predict the frequency up-conversion attainable from a given SL. For the SL considered here, $\omega_R$ values obtained from the equation [open circles in Fig. \ref{fig:spec}] agree well with the numerically-calculated spectrum.  

When $U$ exceeds $U_c$, the bandwidth of $S(\omega)$ increases and the peaks become denser. The spectrum is broad because the Bloch frequency changes continuously throughout each burst. The strongest peaks occur near the maximum frequency of the Bloch oscillations, $\omega_B^{max} =k_SUd/ \hslash$ \cite{esaki-tsu,dohler,WAC02}, whose values are marked by the squares in Fig. \ref{fig:spec}. The series of abrupt jumps [arrowed in Fig. 2(c)] between Bloch oscillation bursts generates the low-frequency ($\omega\lesssim 0.3\times10^{13}$ rad s$^{-1}$) peaks in $S(\omega)$ and their harmonics. These peaks shift abruptly to lower $\omega$ as $U$ increases (most easily seen for $\omega \lesssim 10^{13}$ rad s$^{-1}$). Two distinct series of jumps, each with a period of $\approx1.7$ meV, occur at $U$ values marked by the solid and dashed white lines in Fig. \ref{fig:spec}. Their origin can be understood by considering Fig. 2(d). As $U$ increases, the upper dotted line moves upwards through the $E'(p_x)$ curve. At the $U$ values marked by the solid white lines in Fig. \ref{fig:spec}, the upper dotted line in Fig. 2(d) passes above a local maximum in $E'(p_x)$. This enables the electron to enter a new minizone, so adding an additional Bloch oscillation to each burst [within brackets in Fig. 2(c)]. As a result, the repeat frequency of the bursts decreases abruptly, thus red-shifting the corresponding spectral peaks in $S(\omega)$. Similar shifts occur at $U$ values marked by the dashed white lines in Fig. \ref{fig:spec}, when the lower dotted line in Fig. 2(d) passes below a local minimum in $E'(p_x)$.

In the limit $U\rightarrow0$, $\omega_R\propto U^{1/2}\rightarrow0$, which means that the electron scatters when $x'\approx0$ [Fig. 2(b) inset] and so experiences an almost constant force, $k_SU$. Consequently, in Fig. 1, the $v_d$ curve for the acoustic wave (solid) converges to the Esaki-Tsu curve (dashed) for an electron accelerated by a constant electric field, $k_SU/e$. The two curves also converge when $U \gg U_c$ because the electron immediately experiences a large positive force, equal to the maximum gradient of $V(x')$,which creates Bloch oscillations localised within a distance $\Delta/F$ $(\ll d)$ of $x'=0$. Since $U_c>U_{ET}$ (Fig. 1), the $v_d$ curve produced by the acoustic wave overshoots that generated by a static force and so causes a far higher maximal NDV value, $D_V$. Our analysis predicts strong acoustic enhancement of the peak $v_d$ value, $v_d^{max}$, for all SLs with $\omega_R \tau \approx 1$ when $U \approx U_c$. This ensures that $v_d^{max}$ is close to the mean speed ($\approx \alpha v_S$) of an electron traversing one well in the acoustic wave [Fig. 2(b) inset], rather than the lower speed, $v_S$, of the well itself.

\begin{figure}
  \centering
  \includegraphics*[width=1.\linewidth]{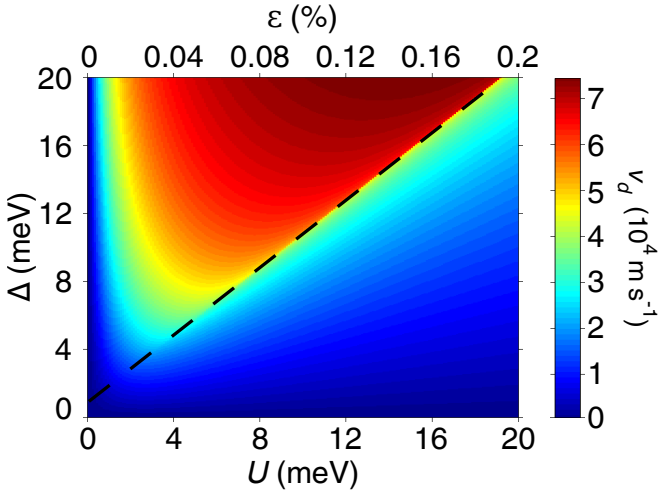}
  \caption{(Color online) Color map of $v_d$ versus $U$ (or $\varepsilon$: top scale) and $\Delta$. Dashed line: $U_c$ versus $\Delta$. $\omega_S=4\times10^{11}$ rad s$^{-1}$.
\label{fig:vdsurface}}
\end{figure}

Figure \ref{fig:vdsurface} shows a color map of $v_d$ calculated versus $U$ (or $\varepsilon$) and $\Delta$. For $U \gtrsim U_c$ (dashed line), $v_d$ decreases abruptly due to the sudden onset of Bloch oscillations. Figure \ref{fig:vdsurface} reveals that the velocity overshoot and, hence, $D_V$, both increase with increasing $\Delta$. When $\Delta$ = 20 meV, $D_V\approx 60$ times higher than for a static force, suggesting that wide miniband, acoustically-driven, SLs will exhibit very high frequency electron dynamics. 

\subsection{Electron dynamics for initial position $x_0\neq0$}

Since the propagating acoustic wave produces a spatially-varying potential it is important to consider how the electron's initial position, $x_0$, affects its subsequent motion. 

\begin{figure}
 \centering
 \includegraphics*[width=1.\linewidth]{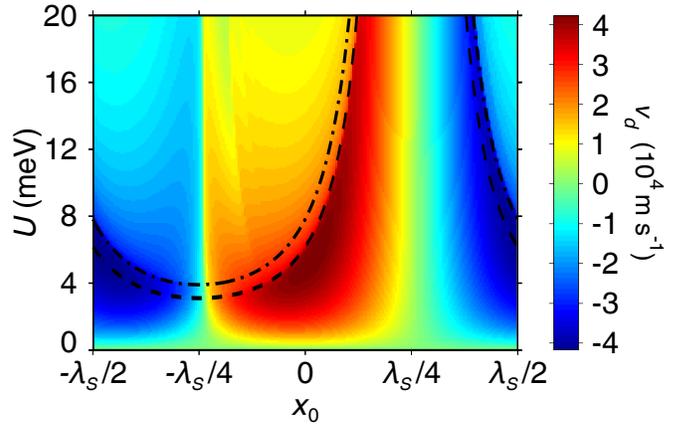}
  \caption{(Color online) Color map showing the electron drift velocity, $v_d$, (scale right) calculated as a function of $x_0$ and $U$. Dashed curve shows $U_c(x_0)$ obtained from Eq. (\ref{eq:Ucphi}) and the dot-dashed curve is $U_c'(x_0)$ calculated using (\ref{eq:Ucphidash}).  
\label{fig:Ucx0_t}}
\end{figure}

Figure \ref{fig:Ucx0_t}, shows $v_d$ calculated as a function of $U$ and $x_0$.  
Due to the spatial periodicity of the propagating wave, $v_d(x_0,U)$ is also a periodic function of $x_0$ with a period equal to $\lambda_S$.

When $x_0=0$, corresponding to the $v_d(U)$ curve shown solid in Fig. \ref{fig:vdenergy}, the color map in Fig. \ref{fig:Ucx0_t} confirms that $v_d$ initially increases with increasing $U>0$. However, when $U=U_c(x_0=0)$ (vertical dashed line in Fig. \ref{fig:vdenergy}), the transition from the wave dragging to the Bloch regime produces a sharp suppression of $v_d$ (red to yellow regions in Fig. \ref{fig:Ucx0_t}). Figure \ref{fig:Ucx0_t} reveals similar behavior for $-\lambda_S/4 \lesssim x_0 \lesssim \lambda_S/8$.
By including $x_0$ explicitly in the Hamiltonian for the system, we find that $H'(x_0) =- U \sin \left(k_S x_0\right)$, meaning that $E'(p_x)=- U \sin \left(k_S x_0\right)-V(x',x_0)$ can only take values between $\pm U - U \sin \left(k_S x_0\right)$. Previously, we showed that for the electron to Bloch oscillate it must attain enough kinetic energy to traverse the first local maximum in the $E'(p_x)$ curve [marked by the arrow in Fig. \ref{fig:trajs}(b)], which occurs when $E'(p_x)\approx \Delta-v_S\hbar \pi /d$ [see Eq. (\ref{eq:uc})].   
Therefore, 
\begin{equation}
 U_c(x_0) \approx  \frac{1}{1-\sin(k_S x_0)}\left(\Delta - \frac{v_S \hbar \pi}{d}\right).
\label{eq:Ucphi}
\end{equation}
Hence, for $U<U_c(x_0)$, the electron is dragged by the acoustic wave, whereas for $U\geq U_c(x_0)$ it is allowed to perform Bloch-like oscillations. The values of $U_c(x_0)$ obtained from Eq.(\ref{eq:Ucphi}) are shown by the dashed curve in Fig. \ref{fig:Ucx0_t}. This curve is in a good agreement with the sudden suppression of $v_d$ in the color map, which results from the onset of Bloch oscillations.

For $\lambda_S/8 \lesssim x_0 \lesssim 3\lambda_S/8$, the transition from the wave dragging to the Bloch oscillation regime occurs beyond the experimentally attainable range of $U$ and there is no associated suppression of $v_d$. Instead, for given $x_0$ between $\lambda_S/8$ and $3\lambda_S/8$, $v_d$ is almost independent of $U$. However, for fixed $U$, increasing $x_0$ produces a gradual decrease in $v_d$. This can be understood  by considering the range of possible values of $E'(p_x)$, given by  $\pm U - U \sin \left(k_S x_0\right)$, which implies that increasing $x_0$ from 0 to $\lambda_S/4$ decreases the maximum attainable value of $E'(p_x)$ [upper horizontal dotted line in Fig. \ref{fig:trajs}(b)]. As a result, the electron can no longer access the high gradient regions of $E'(p_x)$ where the magnitude of $v_x'=dE'(p_x)/dp_x$ is high.  Further increasing $x_0$ causes $v_x'$ to decrease until $x_0 \approx \lambda_S/4$ at which point $E'(p_x)=0$. Consequently, the electron cannot oscillate within a potential well of the acoustic wave and is simply dragged through the lattice at a constant speed $v_S$, so that $v_d\approx v_S$.  

For $x_0> \lambda_S/4$, the initial force on the electron 
\begin{equation}
 -\frac{dV(t=0)}{dx} = k_S U \cos(k_S x_0),
\label{eq:initialforce}
\end{equation}
becomes negative. The electron therefore initially moves in the negative $p_x$ direction where the gradient of $E'(p_x)$, and hence also $v_x'$, is negative [see Fig. \ref{fig:trajs}(b)]. Therefore, $v_d$ becomes increasingly negative as the electron starts to access the high (negative) gradient regions of $E'(p_x)$. 

Figure \ref{fig:Ucx0_t} shows that when $x_0\approx\lambda_S/2$, increasing $U$ from 0 initially reduces $v_d$, increasing its \emph{magnitude}, $|v_d|$. 
However, at a critical value of $U=U_c'(x_0)$, close to $U_c(x_0)$, the \emph{magnitude} of $v_d<0$ decreases dramatically (color map changes abruptly from dark to light blue in Fig. \ref{fig:Ucx0_t}). One might expect that this suppression of $|v_d|$ would occur {\it exactly} at the transition from wave dragging to Bloch trajectories, as seen when $|x_0|<\lambda_S/4$ where the initial force on the electron [Eq. (\ref{eq:initialforce})] is positive. However, for $x_0\approx\lambda_S/2$, the electron is initially forced in the negative $p_x$ direction [see Eq. (\ref{eq:initialforce})]. Therefore, for  $U\approx U_c(x_0)$, the electron scatters before it can traverse the local maximum in $E'(p_x)$ [arrowed in Fig. \ref{fig:trajs}(b)]. Consequently, when $x_0\approx\lambda_S/2$, the transition to the Bloch regime at $U=U_c(x_0)$ has no effect on $v_d$. Instead the suppression of $|v_d|$ occurs when $U$ is slightly larger than $U_c(x_0)$. Specifically, the supression of $|v_d|$ occurs when the electron can Bragg reflect by traversing the local maximum to the \emph{left} of the origin at $p_x=0$, i.e. when $E'(p_x)=\Delta + v_S \hbar \pi/d$. Using this condition in $H'$, we find that 

\begin{equation}
 U_c'(x_0) \approx  \frac{1}{1-\sin(k_S x_0)}\left(\Delta + \frac{v_S \hbar \pi}{d}\right).
\label{eq:Ucphidash}
\end{equation}

The values of $U_c'(x_0)$ obtained from Eq. (\ref{eq:Ucphidash}), shown by the dot-dashed curve in Fig. \ref{fig:Ucx0_t}, coincide almost exactly with the dramatic supression of $|v_d|$ observed when $x_0=\lambda_S/2$. More generally, when $-\lambda_S/4 \lesssim x_0 \lesssim \lambda_S/4$, the initial force is positive [see Eq. (\ref{eq:initialforce})], and $U_c(x_0)$ accurately estimates the $U$ value at which $|v_d|$ is suppressed. However, when $\lambda_S/4 \lesssim x_0 \lesssim \lambda_S/2$ and $-\lambda_S/2 \lesssim x_0 \lesssim -\lambda_S/4$, the initial force is negative [see Eq. (\ref{eq:initialforce})], and $U_c'(x_0)$ gives a better estimate of the position of $|v_d|$ supression. Note that when 
$x_0\approx\lambda_S/4$, so that the initial force is $0$ and $|v_d|$ is minimal, $U_c$=$U_c'\rightarrow \infty$, meaning that the electrons \emph{never} perform Bloch oscillations.


\section{Collective electron dynamics and charge domains}
\label{sec:Sec2} 

To investigate how the acoustic wave affects the collective behavior of the electrons, we solved the current-continuity and Poisson equations self-consistently throughout the device. To do this, we adapted the widely-used drift diffusion model of miniband transport in SLs \cite{schom,FRO04,GRE2009}, for the case of acoustic wave driving. In this model, we discretise the SL region into $N=480$ layers, each of width $\Delta x=L/N=0.24$ nm, small enough to approximate a continuum. The volume electron density in the $m^{th}$ layer (with right-hand edge at $x=m \Delta x$) is $n_m$ and the electric field, $F$, values at the left- and right-hand edges of this layer are $F_m$ and $F_{m+1}$ respectively. In the emitter and collector ohmic contacts, $F= F_0$. The evolution of the charge density in each layer is given by the current continuity equation
\begin{equation}
 e \Delta x \frac{d n_m}{d t} = J_{m - 1} - J_m, \ \ \ m = 1 \ldots N,	\label{eq:continuity} 
\end{equation}
\noindent
where the areal current density from the $m^{th}$ to the $m+1^{th}$ layer is
\begin{equation}
 J_m = e n_m v_{d}^m
- e D \frac{\partial n_m}{\partial x}, \ \ \  m = 1 \ldots N.
\label{eq:currentdensity_diffuion}
\end{equation}

In Eq. (\ref{eq:currentdensity_diffuion}), the drift velocity in the $m^{th}$ layer, $v_d^m$, 
is determined from Eq. (\ref{eq:driftvelocity}), using the semiclassical equations of motion (\ref{eq:simx}) and (\ref{eq:pxdotdef}), in which $x_0=m \Delta x$ is the initial position of the electron, and the Hamiltonian $H(x,p_x)=E(p_x)+V(x,t)-e\overline{F_m}x$ where $\overline{F_m}$ 
is the mean electric field in the $m^{th}$ layer. The diffusion coefficient, $D_E$, is calculated from the Einstein relation $D_E=(k_B T/e)\mu$, 
where $\mu$ is the electron mobility in the linear part of the $v_d(U)$ curve (see Fig. \ref{fig:vdenergy}) \cite{schom} and the temperature $T=4.2$ K. Since $J_m$ depends on the local drift velocity, $v_d^m(V,\overline{F_m})$, the collective electron dynamics depend directly on the single electron orbits. Conversely, the single-particle electron trajectories depend explicitly on the collective electron dynamics through their effect on the charge distribution and, hence, the local electric field $\overline{F_m}$. Our calculations include this mutual dependence because they are based on self-consistent calculations of $\overline{F_m}$ and of the single-particle electron trajectories. 

In each layer, $F_m$ obeys the discretised Poisson equation 
\begin{equation}
F_{m + 1} = \frac{e \Delta x}{\epsilon_0 \epsilon_r} \left( n_m - n_D \right) + F_m, \ \ \ m = 1 \ldots N,  \label{eq:poisson} \\
\end{equation}
\noindent
where $\epsilon_0$ and $\epsilon_r=12.5$ are, respectively, the absolute and relative permittivities and $n_D=3\times 10^{22}$ m$^{-3}$ is the n-type doping density in the SL layers \cite{FRO04}.

We use ohmic boundary conditions, taken from a previous experiment \cite{FRO04,GRE2009}, to determine the current, $J_0 = \sigma F_0$, in the heavily-doped emitter of electrical conductivity $\sigma= $ 3788 S m$^{-1}$. The voltage (=0) across the device is a global constraint, which requires that $0 = V_C + \frac{\Delta x}{2} \sum_{m = 1}^N (F_m + F_{m + 1})$, where the voltage, $V_C$, dropped across the contacts includes the effect of charge accumulation and depletion in the emitter and collector regions and a $17$ $\Omega$ contact resistance \cite{FRO04}. Then, the total current flowing through the SL layers is $I(t)=\frac{A}{N+1}\sum^N_{m=0}J_m$, where $A=5\times10^{-10}$ m$^{2}$ is the cross-sectional area of the SL \cite{WAC02,FRO04,GRE2009}.

\begin{figure}
 \centering
 \includegraphics*[width=1.\linewidth]{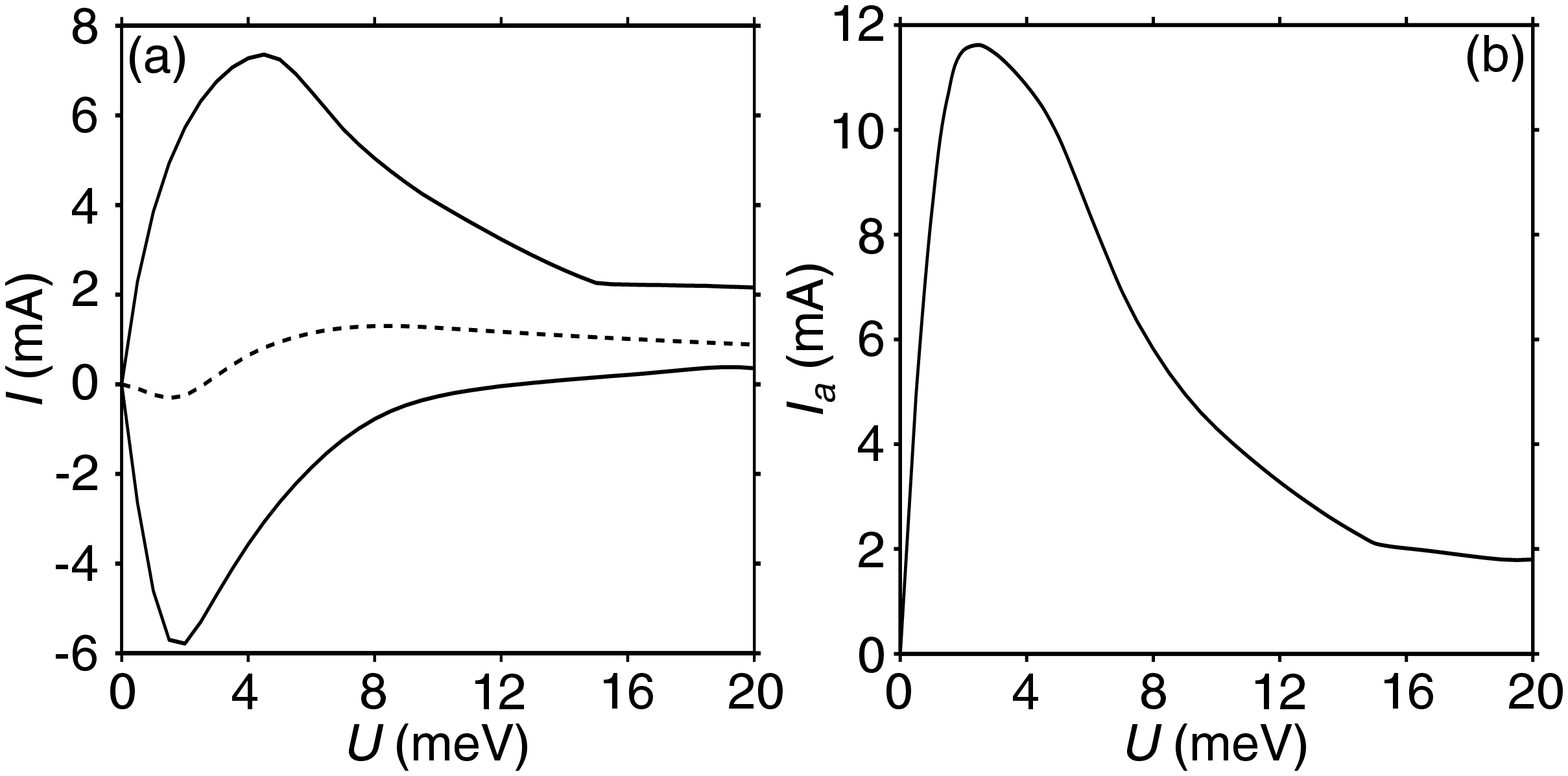}
  \caption{(a) Upper (lower) solid curve: $I_{max}$ $(I_{min})$ calculated as a function of $U$. Dashed curve: time-averaged current, $I_{ave}$. (b) $I_a=I_{max}-I_{min}$ versus $U$.
\label{fig:IU_ink}}
\end{figure}

$I(t)$ oscillates between minimum and maximum values, $I_{\min}$ and $I_{\max}$ respectively, whose variation with $U$ is shown by the upper and lower solid curves in Fig. \ref{fig:IU_ink}(a). The dashed curve in this figure shows the time-averaged current, $I_{ave}$.
Remarkably, even at very low $U$ values the current oscillates. This contrasts with the behavior of SLs with no acoustic wave driving, which exhibit current oscillations only when an applied bias voltage is large enough to induce NDV. Figure \ref{fig:IU_ink}(a) reveals that as $U$ increases from 0 to 4 meV, the magnitudes of $I_{max}$ and $I_{min}$ initially increase, with $I_{max}$ attaining a maximum value of $\approx 7.4$ mA when $U \approx 4$ meV, and $I_{min}$ having a minimum value of $\approx -5.8$ mA when $U \approx 2$ meV. The magnitudes of both $I_{max}$ and $I_{min}$ decrease as $U$ increases beyond $\approx 4$ meV.

In Fig. \ref{fig:IU_ink}(b), we show how the amplitude of current oscillations, $I_a=I_{max}-I_{min}$, changes with $U$. 
Initially, $I_a$ increases with increasing $U$ until it reaches a maximum of $\approx 11.6$ mA when $U \approx 3$ meV. Thereafter, $I_a$ decreases with increasing $U$. The $I(U)$ characteristics shown in Fig. \ref{fig:IU_ink} can be understood within a single-electron picture. As discussed above, and shown in Fig. \ref{fig:Ucx0_t}, all electrons follow dragged orbits when $U\lesssim U_c(x_0=-\lambda_S/4)\approx 3.1$ meV. Consequently, increasing $U$ within this regime raises $v_d$ (see Fig. \ref{fig:Ucx0_t}), thereby also increasing the magnitude of the current oscillations, since $I\propto v_d$. 
As $U$ increases beyond $U_c$, Bloch oscillations gradually turn on, initially for electron trajectories starting at the maximum of the acoustic wave potential (with $x_0 = -\lambda/4$) but eventually for the majority of electrons, thus suppressing the current oscillations.   
\begin{figure}
 \centering
 \includegraphics*[width=.55\linewidth]{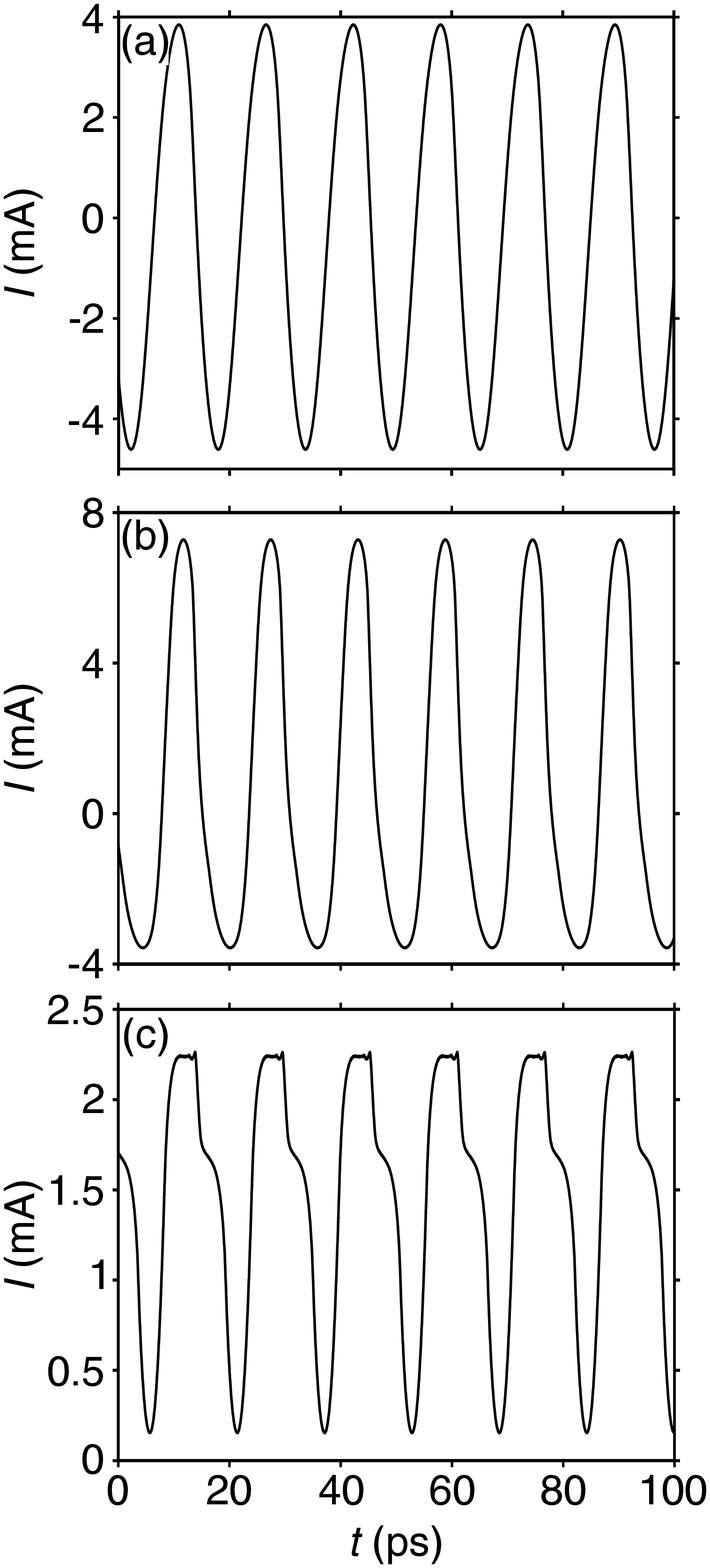}
  \caption{$I(t)$ curves calculated for $U=$ (a) $1$ meV, (b) $4$ meV, (c) $15$ meV.
\label{fig:Itplot_ac}}
\end{figure}

Figure \ref{fig:Itplot_ac} shows $I(t)$ oscillations calculated for three acoustic wave amplitudes: (a) $U=1$ meV, which corresponds to the ascending part of the $I_a(U)$ curve in  Fig. \ref{fig:IU_ink}(b); (b) $U=4$ meV $\approx U_c$, which corresponds to maximal $I_a$ in  Fig. \ref{fig:IU_ink}(b); (c) $U=15$ meV, when $I_a$ is small. Remarkably, the frequency of the $I(t)$ oscillations is independent of $U$ and equals the frequency of the acoustic wave $\omega_S=4 \times 10^{11}$ \rads ($\approx63.7$ GHz).

Figure \ref{fig:Itplot_ac}(a) shows that for small $U$, the $I(t)$ oscillations are almost sinusoidal indicating a single dominant harmonic component. However, as $U$ increases the anharmonicity of the $I(t)$ oscillations also increases [see Fig. \ref{fig:Itplot_ac}(b)]. For large $U$, above $U_c$, the current oscillations are strongly anharmonic [Fig. \ref{fig:Itplot_ac}(c)] due to the appearance of kinks in the $I(t)$ profile.

\begin{figure}
 \centering
 \includegraphics*[width=1.\linewidth]{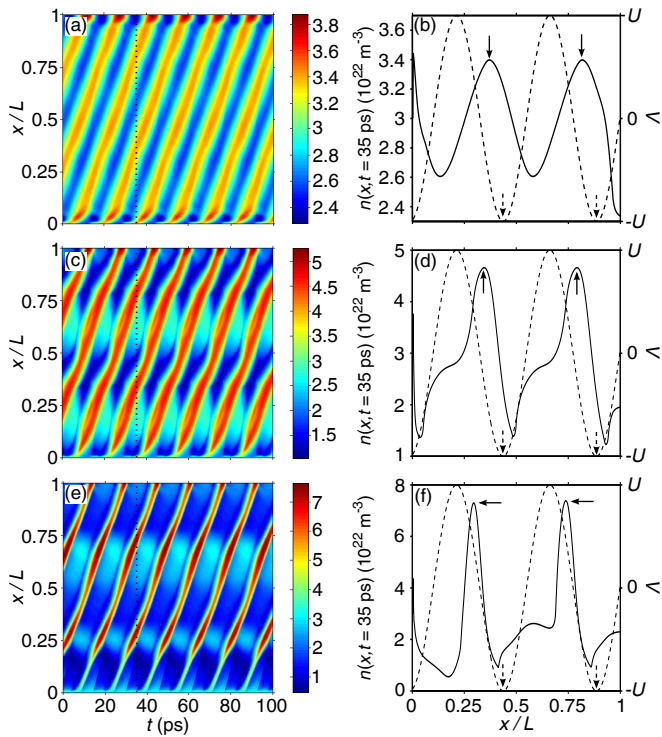}
  \caption{(Color online) Color maps of $n(t,x)$ (left-hand column) and instantaneous electron density profiles, $n(x,t=35$ ps), (solid curves in right-hand column) calculated for $U=$ (a,b) 1 meV, (c,d) 4 meV, (e,f) 15 meV. Scale bars for color map are in units of $10^{22}$ m$^{-3}$. Vertical dotted lines in (a),(c) and (e) indicate time $t=35$ ps corresponding to plots in right-hand column. In (b),(d) and (f), dashed curves show the acoustic wave profile, $V(x)$, at $t=35$ ps (scale on right-hand axes) and solid (dashed) arrows mark electron density maxima (acoustic potential well minima).
\label{fig:chargedomain_surface}}
\end{figure}

To understand the shape of the $I(t)$ oscillations for different $U$, we now examine the spatio-temporal evolution of the electron density, $n(x,t)$, in the SL. Figure \ref{fig:chargedomain_surface}(a) shows $n(t,x)$ calculated for $U=1$ meV within the wave dragging regime. The acoustic wave produces charge density waves, which propagate through the SL with an approximately constant speed ($\approx v_S$) and spatial profile. These charge density waves are a linear response of the SL to perturbation by the plane acoustic wave. In Figure \ref{fig:chargedomain_surface}(b), the solid curve (scale on left-hand axis) shows the spatial form of the charge density wave, $n(t=35$ ps, $x$), i.e. along the vertical dotted line in Fig. \ref{fig:chargedomain_surface}(a). The dashed curve in Fig. \ref{fig:chargedomain_surface}(b) shows the potential energy profile, $V(x,t=35$ ps), of the acoustic wave. Minima in the acoustic wave energy (dashed arrows in Fig. \ref{fig:chargedomain_surface}(b)) lead to the local accumulation of electrons whose density is maximal [solid arrows in Fig. \ref{fig:chargedomain_surface}(b)] near the acoustic wave minima and minimal near the acoustic wave maxima. 
We find that the electron accumulation regions lag slightly behind the local minima in $V$: this is due to inertia as the electrons `ride' up the left-hand sides of the potential wells as the acoustic wave propagates through the lattice from left to right. When each electron accumulation region reaches the collector contact ($x=L$) it produces a sharp increase in $I(t)$. Another period of the charge density wave then forms near the emitter contact and the propagation process repeats, so producing $I(t)$ oscillations \cite{WAC02,GRE2009}. For small $U$, the linear response of the electron gas to the acoustic wave means that the electron accumulation and depletion regions have similar spatial forms and magnitudes, making the $I(t)$ oscillations almost symmetrical around $I=0$ [Fig. \ref{fig:Itplot_ac}(a)].

In the wave dragging regime, increasing $U$ to 4 meV, increases both $v_d$ and $I_a$ [see Fig. \ref{fig:Itplot_ac}(b)]. However, when the wave dragging force is combined with the electric field associated with the charge density modulation, electrons are occasionally driven into the Bloch oscillation regime, which localises them spatially. This localisation induces additional electron accumulation regions [light gray (cyan online) areas in Fig. \ref{fig:chargedomain_surface}(c)], which are known as charge domains and are a nonlinear response of the electron gas to the driving forces. The charge domains appear as plateaux-like features in the (solid) $n(t=35$ ps, $x$) curve in Fig. \ref{fig:chargedomain_surface}(d). This curve is less sinusoidal, and has higher peak values, than for $U = 1$ meV [Fig. \ref{fig:chargedomain_surface}(b)]. Consequently, the $I(t)$ oscillations for $U = 4$ meV [Fig. \ref{fig:Itplot_ac}(b)] are both stronger and more anharmonic than when $U = 1$ meV [Fig. \ref{fig:Itplot_ac}(a)]. 


Further increasing $U$ to 15 meV generally decreases $v_d$ since the electrons are regularly driven into the Bloch oscillation regime. The onset of Bloch oscillations causes electrons to localise and accumulate in certain regions of the SL, thereby creating the high electron density domains shown by the narrow dark gray (dark red online) stripes in Fig. \ref{fig:chargedomain_surface}(e). The associated electric field produces additional isolated islands of electron accumulation, shown by the light gray (cyan online) areas in Fig. \ref{fig:chargedomain_surface}(e). In the (solid) $n(t=35$ ps, $x$) curve shown in Fig. \ref{fig:chargedomain_surface}(f), these additional accumulation regions appear as small broad peaks, which separate the sharper, dominant, maxima. 

Due to the co-existence of the distinct electron accumulation regions in different parts of the SL when $U=15$ meV [i.e. the small and large peaks in Fig. \ref{fig:chargedomain_surface}(f)], the $I(t)$ oscillations [Fig. \ref{fig:Itplot_ac}(c)] have a complex anharmonic form with pronounced kinks just above $I=1.5$ mA. Since the electrons are more strongly localised when $U=15$ meV, than when $U=4$ meV, far fewer electrons per unit time arrive at the collector. Consequently, $I_{max}$ and $I_a$ are both smaller at the higher $U$ value [see Figs. \ref{fig:IU_ink} and \ref{fig:Itplot_ac}(b,c)]. 

Figures \ref{fig:chargedomain_surface}(a),(c), and (e) reveal that a new charge accumulation front forms shortly after each acoustic wave minimum arrives at the left-hand edge of the SL ($x=0$), which occurs with a frequency $\omega_S$. Since this charge subsequently travels through the SL, $I(t)$ oscillates periodically at a frequency $\omega_S$ independent of the value of $U$.

Note that the electric field resulting from electron redistribution within the SL perturbs the single-electron trajectories, but has no qualitative effect on the crossover between the wave dragging [Fig. \ref{fig:chargedomain_surface}(a)] and Bloch oscillation [Fig. \ref{fig:chargedomain_surface}(c)] regimes. However, the transition between these regimes [Fig. \ref{fig:chargedomain_surface}(b)] is blurred slightly by the local electric field. This is because, near the crossover, in parts of the SL the local electric field opposes the electron acceleration due to the acoustic wave, thus maintaining wave dragging. But in other regions, the electric field supplements acceleration by the acoustic wave, so driving the electrons further into the Bloch oscillation regime.

\section{Conclusion}
\label{sec:Sec3}

In conclusion, acoustic waves can induce an abrupt transition between two distinct dynamical regimes of electron transport in SLs. When $U<U_c$, the electrons oscillate within a single spatial period of the acoustic wave, at frequencies $\omega_R \gg \omega_S$, and are dragged through the SL with a drift velocity whose peak value can greatly overshoot that produced by a \emph{static} field. For $U>U_c$, the acoustic wave triggers bursts of Bloch oscillations, thus causing very high NDV. 

The acoustic wave causes the formation of charge density waves and, when $U>U_c$, nonlinear charge domains, which propagate through the SL and thereby create self-sustained current oscillations. The fundamental frequency of these oscillations equals that of the acoustic wave for all $U$. However, both the shape of the $I(t)$ curve, and the spatial profile of the charge domains, depend on $U$ in a way that reflects the underlying single-particle dynamics. In the wave-dragging single-particle regime ($U < U_c$), increasing $U$ strengthens the electron accumulation and depletion in the charge density waves, and hence increases the magnitudes of $I_{max}$, $I_{min}$ and $I_a$. Conversely, when $U > U_c$, the onset of Bloch oscillations creates additional (nonlinear) charge domains, which co-exist with the (linear) charge density waves shaped by the acoustic wave period. In this regime, as $U$ increases the electrons become increasingly localised, reducing the magnitudes of $I_{max}$, $I_{min}$ and $I_a$.

The complex single and collective electron dynamics that we have identified create new perspectives for using acoustic waves to generate high-frequency electric current oscillations. They also highlight the potential of SLs to bridge the interface between conventional electronics and the rapidly-developing field of phononics \cite{CHA06,WANG07}. For example, SLs could be used to transform acoustic waves into electro-magnetic ones, produced by the formation and propagation of charge domains, and, conversely, to detect propagating acoustic waves via the current oscillations that they induce. 

Finally, we note that cold atoms in optical lattices may exhibit similar dynamics \cite{renzoni,Scott02}, with the abrupt transition to Bloch oscillations providing sensitive control of transport.

\section{Acknowledgements}
The work was supported by the UK Engineering and Physical Sciences Research Council. AB acknowledges financial support from the Federal Agency for Science and Innovation of Russia. We acknowledge helpful discussions with E.V. Ferapontov (Loughborough University).


\begin{thebibliography}{9}
\bibitem{foot2}Such a wave can be generated by using `SASER' SLs \cite{kent_saser} or ultra-fast excitation of a thin metal film \cite{kent_metalfilm}.
\bibitem{footac}Semiclassical electron trajectories in the miniband have a fundamentally different character from classical orbits in a single quantum well with the same applied fields: compare, for example, \cite{FRO04} with \cite{FRO94,FRO97}.
\bibitem{footnoteD}The electron dynamics are insensitive to $k_S$ provided, as here, $k_S \ll 2\pi/d$ so that $\lambda_S \gg d$ and $v_S$ is constant.
\end{thebibliography}
\end{document}